\documentclass[a4paper,12pt]{article}

\usepackage{amsfonts}
\usepackage{amssymb}
\usepackage{epsfig}

\def\beq{\begin{equation}}
\def\eeq{\end{equation}}
\def\bea{\begin{eqnarray}}
\def\eea{\end{eqnarray}}
\def\winf{W_{1+\infty}\ }
\def\u1{\widehat{U(1)}}
\def\su2{\widehat{SU(2)}_1}

\def\I{{\rm Im}\ }

\def\nl{\nonumber \\}

\def\a{\alpha}
\def\b{\beta}
\def\g{\gamma}

\def\k{\kappa}
\def\l{\lambda}
\def\t{\tau}
\def\de{\partial}
\def\eps{\varepsilon}
\def\ov{\overline}

\catcode`\@=11
\def\numberbysection{\@addtoreset{equation}{section}
        \def\theequation{\thesection.\arabic{equation}}}
\numberbysection

\begin{document}

\begin{titlepage}
\begin{center}
\hfill  \quad DFF 375/7/2001 \\
\hfill  \quad cond-mat/0111437 \\
\vskip .5 in
{\Large\bf Thermal Transport}

{\Large\bf in Chiral Conformal Theories}

{\Large\bf and Hierarchical Quantum Hall States}
\vskip 0.3in

Andrea CAPPELLI\\
{\em I.N.F.N. and Dipartimento di Fisica, Via G. Sansone 1, I-50019
Sesto Fiorentino (FI), Italy}
\\
\vskip 0.3in
Marina HUERTA\\
{\em Centre de Physique Theorique- C.N.R.S.,
Campus de Luminy, Case 907, F-13288 Marseille, France}
\
\vskip 0.3in
Guillermo~R.~ZEMBA\footnote{
Fellow of CONICET, Argentina.}
\\
{\em Physics Department, C.N.E.A.
% Comisi\'on Nacional de Energ\'{\i}a At\'omica, 
Av.Libertador 8250, (1429) Buenos Aires, and
Universidad Favaloro, Sol{\'{\i}}s 453, (1078) Buenos Aires, 
Argentina}
\end{center}
\vskip .5 in
\begin{abstract}
Chiral conformal field theories are characterized by a ground-state
current at finite temperature, that could be observed, e.g. 
in the edge excitations of the quantum Hall effect.
We show that the corresponding thermal conductance is 
directly proportional to the gravitational anomaly
of the conformal theory, upon extending
the well-known relation between specific heat and conformal anomaly.
The thermal current could signal the elusive neutral edge modes
that are expected in the hierarchical Hall states.
We then compute the thermal conductance for 
the Abelian multi-component theory and the $W_{1+\infty}$ minimal model,
two conformal theories
that are good candidates for describing the hierarchical states.
Their conductances agree to leading order but differ in the 
first, universal finite-size correction, that could
be used as a selective experimental signature.
\end{abstract}

%\bigskip


\vfill
\end{titlepage}
\pagenumbering{arabic}

%-1------------------------------------

\section{Introduction}

The Jain hierarchical states \cite{jain} are an interesting class of
plateaus in the fractional quantum Hall effect \cite{prange}
that still requires further theoretical and experimental investigation.
A characteristic new feature with respect to the Laughlin's plateaus, 
is the presence of edge excitations with neutral
propagating modes that cannot be directly measured in 
the conduction experiments.
In this paper, we elaborate the proposal of Ref.\cite{kafi} of testing 
these neutral excitations by means of their thermal current.

We first use the conformal field theory (CFT) description of the
edge states to obtain a general relation between
the thermal conductance and the gravitational anomaly \cite{anom}
of the chiral conformal theory.
We start by recalling the well-known CFT result expressing 
the specific heat in terms of the conformal anomaly $c$ \cite{affl}:
\beq
c_V= \frac{\partial\langle{\cal E}\rangle_T}{\partial T}
= \frac{\pi\ k_B^2 T}{3v}\ c \ ,\qquad\qquad T \to 0\ .
\label{gen-cv}
\eeq
The proof of this formula goes as follows: the thermal field theory 
is defined on the cylinder geometry made by the periodic Euclidean time
and the unbounded space; in this geometry, the Casimir effect
implies a non-vanishing ground-state energy 
$\langle{\cal E}\rangle_T$, that is given by the expectation value
of the stress tensor and is proportional 
to the conformal anomaly $c$ \cite{cft}.

The edge excitations of the quantum Hall effect are described by 
more general CFTs with unbalanced chiral and anti-chiral modes that propagate
in opposite directions and are characterized by different central charges
$c$ and $\bar{c}$, respectively.
In these chiral theories, the expectation values of the energy 
$\langle{\cal E}\rangle_T$ and the momentum $\langle{\cal P}\rangle_T$
are tied together, such that the Casimir effect is associated to a
thermal current $J_Q\propto \langle {\cal P}\rangle_T $.
The corresponding thermal Hall conductance $K$ can be found in full
generality by adapting the derivation of (\ref{gen-cv}) (see Section 2); 
it reads:
\beq
K\equiv \frac{\partial J_Q}{\partial T}= 
\frac{\pi\ k_B^2 T}{6}\ \left(c - \bar{c} \right)\ .
\label{gen-k}
\eeq
In this Equation, the difference of central charges parametrizes
the two-dimensional (pure) gravitational anomaly 
\cite{anom} of the chiral conformal theory describing the single edge (e.g. 
$c=1, \bar{c}=0$ for the edges of the Laughlin plateaus).
Equation (\ref{gen-k}) generalizes the results of Ref.\cite{kafi}.

The formula (\ref{gen-k}) for the thermal conductance
allows for the direct determination of 
the central charges of the edge states; clearly,
the measurement of $K$ requires a high-precision experiment
that has not yet been realized, to our knowledge \cite{expe};
nevertheless, it should be feasible in principle \cite{kafi}.  
On the other hand, the specific heat of edge states (proportional to
$(c+\bar{c})/2$, in general) cannot be measured because it
is masked by the overwhelming contribution of the lattice
phonons \cite{prange}.

In a finite sample, of typical linear extension $R$, 
there are finite-size corrections to the leading ($R\to\infty$)
results (\ref{gen-k}) and (\ref{gen-cv}).
The  exact thermal averages are then obtained 
by differentiation of the partition function;
the latter can be best computed on the geometry of the annulus, 
that involves an inner and an outer edge. 
Actually, the annulus partition functions enjoy the property of 
invariance (or, more 
generally, of covariance) under modular transformations of the
periodic time and angular coordinates \cite{czmi}.
Once the expression of the partition function is known,
one can determine the finite-size corrections to (\ref{gen-k}):
in particular the first, universal, term $O(1/R)$ is of some
interest as described hereafter.

The edge excitations of the hierarchical Hall states
have been mainly described by two classes of conformal field theories:
the Abelian theories \cite{abe}, and the $\winf$ minimal models \cite{ctzmin}. 
Both theories are suitable generalizations of the simple scalar field theory 
that describes Laughlin's plateaus (chiral Luttinger liquid) \cite{wen}. 
The first class considers several copies of the scalar theory, 
whereas the second class exploits the physical picture of the droplet of
incompressible fluid \cite{laugh}, that is
characterized by the $\winf$ symmetry \cite{ctz}\cite{iks}.
Remarkably, these two classes of theories possess the same spectra of
charged excitations, but differ in the neutral sector.

Therefore, the thermal conductances are expected to be different in the two 
theories and could provide a significant experimental test.
Actually, the conductances are equal to leading order (\ref{gen-k}), because
the two theories share the same central charges.
From the known expressions of their partition functions \cite{czmi},
we nevertheless obtain different first-order $O(1/R)$ corrections: 
they are vanishing in the Abelian theories, and non-vanishing in
the $\winf$ minimal models.
The derivation makes it apparent that power-law $O(1/R^k)$ corrections 
are absent for all rational CFT, namely the class of CFTs with modular 
invariant partition functions. 

The outline of the paper is the following.
In Section 2, we describe the thermal transport in CFT and
derive the general formula (\ref{gen-k}); we then digress on
the relation between the anomalies, chiral and gravitational,
and the corresponding out-of-equilibrium processes of the Hall and 
thermal currents, respectively. These are nice examples
for the general picture of non-equilibrium dynamics
envisaged in Ref.\cite{poly}.
In Section 3, we introduce the annulus partition function
and obtain the finite-size corrections to (\ref{gen-k}),
using the well-known $c=1$ CFT of the Laughlin plateaus as an
example; we then extend the finite-size analysis to the
multi-component Abelian CFT of the hierarchical states.
In Section 4, we analyze the $\winf$ minimal model and
find the $O(1/R)$ correction stemming from the lack of
modular invariance of its partition function.
In the Conclusions, we estimate the experimental precision
that is necessary for measuring these finite-size
differences.

%-2.1---------------------------------
\section{Thermal conductance in conformal field theory}

\subsection{Earlier results}

We start by recalling the results of Ref.\cite{kafi}
for the thermal transport of the edge excitations that are
described by $m$ independent $(1+1)$-dimensional scalar fields.
The thermal current $J_Q$ was defined as:
\beq
J_Q=\sum_{i=1}^{m}\ \eta _{i}\ v_{i}\ \varepsilon_{i}\ ,  
\label{thcu}
\eeq
where $v_{i}$ and $\varepsilon _{i}$ are, respectively, the velocity 
and the average energy density of the $i$-th mode, 
and the chiralities $\eta_{i}=\pm 1$ indicate the sense of propagation.
The thermal conductance was given by:
\beq
K\ =\ \frac{\partial J_Q}{\partial T}\ =\ \sum_{i=1}^{m}\ \eta _{i}\
v_{i}\ \frac{\partial \varepsilon _{i}}{\partial T}\ . 
\label{kdef}
\eeq
In Ref.\cite{kafi}, the energy densities $\eps_i$ were 
obtained by assuming linear dispersion relations\footnote{
Hereafter, we set $\hbar=c=1$.}
$E_i(k)=v_i k$, $k>0$ being the momentum of chiral excitations,
and by thermal averaging with independent Bose distribution functions
at temperature $T$. The result was, in the thermodynamic limit,
\beq
\eps_i = \frac{1}{v_i} \frac{\pi\ k_B^2 T^2}{12}\ ,
\label{kf-eps}
\eeq
and led to the thermal conductance:
\beq
K= \frac{\pi\ k_B^2 T}{6}\ \sum_{i=1}^m \eta_i \ .  
\label{kf-k}
\eeq
One interesting remark of Ref.\cite{kafi} was that this result
only depends on the (topological
invariant) sum of chiralities, that can vanish or even be negative
if the edge contains a number of neutral modes propagating 
oppositely to the electric current.
Actually, the multi-component scalar theory \cite{abe} 
for the hierarchical plateaus at filling fraction $\nu=m/(p m +1)$
(resp. $\nu=m/(p m - 1)$), with $m=2,3,\dots$ and $p=2,4,\dots$, predicts
one charged mode and $(m-1)$ neutral ones, whose chiralities
are all equal (resp. all opposite) to that of the charged mode.
For example, the thermal conductance was predicted to vanish
for $\nu=2/3$ and be negative for $3/5$.

%-------Fig 1----------------------------------------------------
\begin{figure}
\epsfysize=4cm
\centerline{\epsfbox{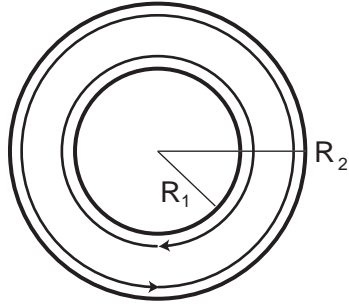}}
\caption{Annulus geometry with inner and outer radii $R_{1}$ and
$R_{2}$, respectively. The arrows indicate the sense of propagation 
of the edge excitations.}
\label{fig1}
\end{figure}
%-------Fig 1----------------------------------------------------

%-2.2-------------------------

\subsection{General formula}

In the following, we would like to generalize the previous results
using the CFT methods \cite{cft}.
In Euclidean coordinates $z=v\tau +i x$,
the local energy ${\cal E}(x,\tau)$ 
and momentum ${\cal P}(x,\tau)$ densities, and their
continuity equation, are expressed in
terms of the components of the stress tensor ${\cal T}(z)$ and 
$\ov{\cal T}(\bar{z})$ as follows:
\bea
&&\frac{\de}{\de \bar{z}}\ {\cal T} = \frac{\de}{\de z} \ \ov{\cal T}=0 \ ,\nl
&&{\cal E} = \frac{v}{2\pi} \left({\cal T} +\ov{\cal T} \right)\ , \nl
&& {\cal P} = \frac{v^2}{2\pi} \left({\cal T} -\ov{\cal T} \right)\ .
\label{j-def2}
\eea
The (steady) thermal current is defined
as the ground-state value of the {\it momentum density} in the 
thermal field theory:
\beq
J_Q \equiv \langle {\cal P}\rangle_T =
\frac{v^2}{2\pi} \left\langle {\cal T} -\ov{\cal T} \right\rangle_T\ ;
\label{j-def}
\eeq
thus, it is proportional to the difference of 
the two chiral components of the stress tensor and is 
non-vanishing for chiral conformal theories\footnote{
Persistent charge currents along the edge have been discussed in
Ref.\cite{ino}.}.

In the quantum Hall effect, this analysis applies to the
single edge, that is typically described by a chiral CFT  
(e.g. $\ov{\cal T}=0$ for the Laughlin states);
nevertheless, any Hall device, like the bar or the annulus 
(Fig.\ref{fig1}),
always involves two edges with conjugate chiralities;
therefore, the total thermal current vanishes in thermal equilibrium.
Let us focus on the annulus geometry, characterized by the 
space-independent currents $J_Q^{(1)}(T)$ and $J_Q^{(2)}(T)$,
with indices numbering the edges, that satisfy 
$J_Q^{(1)}(T)+J_Q^{(2)}(T)=0$.
The non-equilibrium current can be described as follows: 
we homogeneously heat up one of the edges,
say the inner one (1), and find, within the linear response:
\beq
\Delta J_Q =J_Q^{(1)}(T+\Delta T)+J_Q^{(2)}(T) \sim 
\frac{\partial J_Q^{(1)}}{\partial T}\ \Delta T \ .
\label{J-der}
\eeq
Therefore, the thermal Hall conductance is defined by:
\beq
K\equiv \frac{\partial J_Q^{(1)}}{\partial T}= \frac{v^2}{2\pi} 
\frac{\partial}{\partial T}\left\langle {\cal T} 
-\ov{\cal T} \right\rangle^{(1)}_T\ ,
\label{k-der}
\eeq
and it involves the stress-tensor components of the single-edge CFT.
This formula is consistent with the earlier definition
(\ref{kdef}), but it is also an exact result that can be used
for computing the finite-size corrections.
In the annulus geometry, the thermal current is longitudinal, i.e.
 orthogonal to the temperature gradient,
due to the presence of the external magnetic field 
(this thermal analog of the Hall conduction is
called the Leduc-Righi effect in magneto-hydrodynamics \cite{landau}).

We now proceed to compute the thermal conductance (\ref{k-der}) to
the leading $(R\to\infty$) order; disregarding the space periodicity,
the thermal field theory is defined on the cylinder
made by the unbounded space and the periodic Euclidean time,
with period $\b=1/k_B T$.
The thermal average of the stress tensor is obtained by
mapping the cylinder to the two-dimensional Euclidean plane
with the help of the conformal transformation:
\beq
z(w)=\exp\left[ {i2\pi \left(v\tau +ix \right) \over v \b }
\right] \quad\longleftrightarrow\quad w=v\tau +ix \ .
\label{map-t}
\eeq
The transformation law of the stress tensor involves an anomalous term
given by the Schwarzian derivative $\{z,w\}$ times the Virasoro central 
charge \cite{cft}:
\bea
\langle {\cal T}(w)\rangle_{\rm cyl} &=&  
\langle {\cal T}(z)\rangle_{\rm plane}
\left(\frac{dz}{dw}\right)^2 +
\frac{c}{12} \left\{ z,w \right\} \nl
& = & 0+
\frac{c}{12} \left[ \frac{z'''(w)}{z'(w)} -\frac{3}{2}
\left( \frac{z''}{z'} \right)^2 \right] =
\frac{\pi^2 \ c}{6 v^2 \b^2}\ . 
\label{schwa}
\eea
Using Eq.(\ref{j-def}), the thermal current is found to be a
translation-invariant constant, that is equal
to the space average of the momentum density, i.e., the zero mode 
on the cylinder $\left(L_{-1} -\ov{L}_{-1} \right)_{\rm cyl}$:
\beq
J_Q= \frac{v^2}{M}\left(L_{-1} -\ov{L}_{-1} \right)_{\rm cyl}=
\frac{v^2}{M}\int_{-iM/2}^{iM/2}\ \frac{dw}{2\pi i}
\left({\cal T}(w) -\ov{{\cal T}(w)} \right) = \frac{\pi}{12}k_B^2 T^2
\left( c-\bar{c}\right)\ .
\label{gen-e}
\eeq
In this equation, we temporary set a cutoff $M$ in the
unbounded spatial direction along the cylinder axis.
Upon differentiation w.r.t. the temperature, we obtain the 
general formula for the thermal Hall conductance that was anticipated
in the Introduction (\ref{gen-k}):
\beq
K= \frac{\pi\ k_B^2 T}{6}\ \left(c - \bar{c} \right)\ .
\eeq
Some comments are in order: 

i) The Virasoro central charges $(c,\bar{c})$
parametrize both the conformal and gravitational anomalies, which
can be traded into one another by changing the renormalization 
scheme; if $c=\bar{c}$, the gravitational anomaly can be set to
zero, which is the usual case for statistical mechanics models;
if $c\neq\bar{c}$, the minimal gravitational anomaly is
proportional to $(c-\bar{c})$.
The CFT of one Hall edge is necessarily chiral and anomalous.

ii) the corresponding result for the specific heat is:
\beq
c_V =\frac{\partial \langle {\cal E}\rangle_T}{\partial T} =
\frac{\pi\ k_B^2 T}{6v}\ \left(c + \bar{c} \right)\ ,
\label{cv-ii}
\eeq
in agreement with \cite{affl}.
Although this is a well-established result that has been 
widely used in numerical and real experiments in statistical
mechanics, the corresponding result for the thermal conductivity
was not fully appreciated, possibly because chiral CFTs
are rather unusual in this domain.

iii) The general expression
reproduces the earlier result (\ref{kf-k}) for the multi-component
scalar field theory \cite{kafi}, since each field contributes to 
$c$ or $\bar{c}$ by one unit.
Note also that the dependence on the velocity cancels out in the final 
result of the thermal conductance, that is fully universal;
in case of independent velocities for chiral and
antichiral modes, $v$ and $\bar{v}$, respectively, one
should write an independent map to the cylinder for each
chiral part, involving $w$ as above and $\bar{w} =\bar{v}\tau +ix$.

%-2.3----------------------------------

\subsection{Anomalies and non-equilibrium processes}

We have seen that the Hall and thermal currents have some
striking similarities: both are orthogonal to the magnetic field
and the applied force, the in-plane electric field or the
temperature gradient, respectively.
Both currents correspond to an out-of-equilibrium steady
motion that is dissipationless due the orthogonality to the force.
Another fact is that these currents are associated to the two
anomalies of the chiral CFT: the chiral and gravitational ones, respectively.
It is interesting to discuss the physical mechanisms 
underlying the two flows, that are tractable cases of 
non-equilibrium dynamics.

The relation of the Hall current to the chiral anomaly of the edge CFT is 
well understood \cite{cdtz}: the mechanism is that of the ``spectral flow'',
in which the leak of electric charge $Q$ out of (and orthogonal to) the edge
is caused by the electrons that are pulled out of the Dirac sea
by the applied tangential electric field $E$.
Indeed, the chiral anomaly equation:
\beq
\de_{\bar z} J= \frac{\k e^2}{2\pi}\ F\ ,\qquad F_{ij}=\eps_{ij} F=
\de_i A_j -\de_j A_i\ , \quad i=\hat{t},\hat{x},
\eeq
with $x$ the tangential coordinate, can be integrated over one edge
of the annulus,
\beq
\frac{\de Q^{(1)}}{\de t} =\frac{\k e^2}{2\pi} \int dx E_{\hat{x}} \ ,
\eeq
and can be recognized as the radial Hall current 
$J_{\hat{r}} =\sigma_H E_{\hat{x}}$
in $(2+1)$ dimensions; the Hall conductivity, $\sigma_H =e^2 \nu/2\pi$, 
 is parametrized by $\nu=\kappa$, a combination of the charge unit
and the coupling constant of the CFT.
This is clearly a steady out-of-equilibrium process, 
where the charge is not conserved, but the flux of the corresponding
current does. Such states have been called ``flux states'' by Polyakov 
\cite{poly}, which has stressed the relevance of anomalies
in quantum field theory for modeling non-equilibrium processes.
The quantum Hall effect provides two neat examples of this picture,
that are not hampered by the difficulties of describing the dissipation. 

The gravitational anomaly amounts to the non-conservation
of the stress tensor in a gravitational background:
\beq
\nabla^z T_{zz} = - \frac{c}{24} \nabla_z {\cal R}\ ,
\label{gr-an}
\eeq
where $\nabla_z$ is the covariant derivative and ${\cal R}$ the scalar
curvature of the background metric. Upon integration of this equation,
one obtains the anomalous transformation of the stress-tensor
involving the Schwarzian derivative (\ref{schwa}) \cite{cft}, that
yields the Casimir effect and the thermal current.
In less technical terms, the mechanism can be explained as follows:
on the cylinder geometry of the thermal field theory,
the constant, steady thermal flow implies that energy is conserved
locally but not globally; there exist a source and a drain at space
infinity, $x=\pm\infty$, where the curvature is non-vanishing
and the conformal mapping (\ref{map-t}) is singular, in agreement
with the anomaly equation (\ref{gr-an}).
On the annulus, each edge is closed, the points at space infinity are
identified and there is no non-conservation: 
the total thermal current is non-vanishing when the two edges
have different temperatures (different backgrounds).

In the case of turbulence \cite{poly}, the energy flux is constant
in momentum, rather than in space, and the singular points at
infinity correspond to the infrared (resp. ultraviolet) limits,
where energy is injected (resp. dissipated).
It is possible that chiral conformal theories defined on other
gravitational backgrounds may yield further models of 
out-of-equilibrium processes.

%-3.1-----------------------------------------

\section{Annulus partition function and finite-size corrections}

\subsection{The Laughlin states}
 
Now we would like to evaluate the finite-size correction to
the infinite-volume result (\ref{gen-k}); we shall obtain them 
by differentiation of the partition function, that completely accounts
for the properties of the spectrum including the finite-size effects.
As an example, we shall discuss the simplest CFT with central charge $c=1$,
that describes the Laughlin states.

We consider again the annulus geometry (Fig 1), whose
partition function can be computed 
using the data of the representation theory of the Virasoro algebra; 
moreover, this partition function
obeys the powerful constraints of modular invariance (covariance)
\cite{czmi}.
In the course of the argument, it will become apparent that the first
finite-size correction is universal and equal to that of the bar geometry.

The partition function is defined by the trace over the Hilbert
space of the Boltzmann weight, that can be expressed in terms of
the Virasoro generators \cite{cft}: one employs the map from
the $z$-plane to the $u$-cylinder with space period $2\pi R$:
\beq
z =\exp\left( \frac{u}{R} \right)\ \longleftrightarrow \ u=v\tau +ix
\label{r-map}\ .
\eeq
Proceeding as in the previous Section, we find that the total
energy-momentum on the cylinder is transformed as follows \cite{cft}:
\beq 
\left( L_{-1} \right)_{\rm cyl} =
\frac{1}{R} \left[\left( L_0 \right)_{\rm plane} -\frac{c}{24} \right]\ ,
\label{l-1}
\eeq
and similarly for the other chirality.
The eigenvalues of the $L_0, \bar{L}_0$ operators in the plane 
are determined by the representation theory of the relevant
chiral algebras.

The annulus partition function, defined e.g.
in Ref.\cite{czmi}, describes the full physical system involving 
both the inner and outer edges,
labeled by the indices $\kappa =1,2$, and having conjugate chiralities.
The Hamiltonian on one edge, say $(1)$, is:
$H_1=(v_1/R_1)\left( L^{(1)}_{0} -c_1 /24 \right) +
(\bar{v}_1/R_1)\left( \ov{L}^{(1)}_{0} -\bar{c}_1 /24 \right)$;
on the other edge, it is the conjugate expression 
($c_2=\bar{c}_1$, $\bar{c}_2=c_1$) parametrized by $\{v_2,\bar{v}_2,R_2\}$.
It is natural to assume the equilibration between the edges, i.e.
 $v_{1}/R_{1}=v_{2}/R_{2}$.
The full theory is characterized by the total Virasoro operators,
$L_0 =L^{(1)}_{0}+L^{(2)}_{0}$ and its conjugate,
and by the total central charge $c=c_1+c_2=\bar{c}$.
The partition function takes the form:
\beq
Z(\tau,\overline{\tau})\ =\ {\rm Tr}\left[ \ 
q^{L_{0}-c/24}\ {\overline{q}}^{{\overline{L}}_{0}-c/24}\ 
\right] \ ,  
\label{zdef}
\eeq
where the trace extends over all the states in the Hilbert space.
The parameters 
$q=\exp (2\pi i\tau )$ and ${\overline{q}}=\exp (-2\pi i{\overline{\tau }})$,
with $\tau=(v/2\pi R)(\gamma + i \beta) $ and 
${\overline{\tau }} =(\bar{v}/2\pi R)(\gamma-i\beta)$ 
encode the dependence on the temperature,
$\beta =1/k_{B}T$, and the ``torsion'' $\gamma$.
In general, the partition function would also contain
the dependence on the electric potential, conjugate to the charge of the
excitations \cite{czmi}, but this is discarded here.

The edge states of the Laughlin plateaus at $\nu=1/p$, $p=1,3,5,\dots$,
are described by the CFT of
the chiral boson field compactified on a circle of radius
$r^2=1/p$  \cite{wen}\cite{cdtz}; its central charge $c=1$
accounts for a single chiral propagating mode per edge
($c_1=\bar{c}_2=1$, $c_2=\bar{c}_1=0$).
The partition function is given by \cite{czmi}:
\bea
Z(\t,{\overline{\t}}) & =& \sum_{\lambda =1}^{p}\ \chi_\l (\t)\
\overline{\chi_\l (\t)}\ ,\nl
\chi_\l (\t) & =&
\frac{1}{\eta (\t)}\ \sum_{k=-\infty}^\infty\ q^{(pk+\lambda )^{2}/2p}\ ,
\qquad
\eta (\t)=q^{1/24}\prod\limits_{k=1}^{\infty }(1-q^{k})\ , 
\label{zucom}
\eea
where each $\chi _{\lambda }$ is a sum of characters of the Abelian
current algebra $\widehat{U(1)}$ and  $\eta (\t)$ is the 
Dedekind function.

In order to compute the thermal current on one edge, say $(1)$,
we need to split the statistical sum into parts pertaining to each
edge. These are given by the
``chiral partition functions'' $\chi_\l (\t)$ in (\ref{zucom}): 
actually, these sums involve states of a single edge and
depend on the parameter $\l=1,\dots,p$ specifying the sector of
fractional charge $Q=\l/p +{\rm integer}$ (see also Ref.\cite{ino}).
As shown by Eq.(\ref{zucom}), the chiral partition functions on
the two edges are only coupled through the integrality condition on the
total charge.

The (constant) thermal current (\ref{j-def},\ref{gen-e}) can be
rewritten:
\beq
J_Q = v \left(\eps_1 -\ov{\eps}_1 \right)\ ,
\label{jq}
\eeq
in terms of the average chiral energy densities on the first edge; 
in the chiral boson theory, $ \ov{\eps}_1=0$ and 
\beq
\eps_1(\l) = - \frac{i}{2\pi R} \left. 
\frac{\partial \log \chi_\l}{\partial \g} \right\vert_{\g =0} \ ,
\label{eps1}
\eeq
that may depend on the topological sector.
Note that differentiation w.r.t. $\b$ would not take the 
chirality sign of (\ref{jq}) into account.

Next, we evaluate this quantity in the low-temperature, $\b\to\infty$,
and large-size limit, $ v\b \lesssim R$, i.e. small $x=v\b/R$,
keeping the first finite-size correction.
The sum in the numerator of $\chi_\l$ (\ref{zucom}) can be approximated by a
continuous gaussian integral, that is actually $\l$-independent; 
in the denominator, the derivative of the Dedekind function 
is a sum that is also approximated by an integral plus 
the first finite-size correction given by the Euler-Maclaurin formula.
The result is:
\bea
J_Q &=& \frac{v^2}{2\pi R^2} \left[ \left(
\frac{1}{2x} + O(1) \right) + \left( \frac{\pi^2}{6x^2} -
\frac{1}{2x} + O(x) \right)\right] \nl
&=& \frac{\pi}{12 \b^2}
\left( 1+ O\left( \frac{\b^2 v^2}{R^2} \right)\right)\ ,
\qquad\qquad\qquad \left( x=\frac{v\b}{R} \right)\ .
\label{epsc1}
\eea
In the first line of this Equation, 
the first (resp. second) parenthesis contains the
contribution of the numerator (resp. denominator) of $\chi_\l$.
We thus reproduce the general result
Eq. (\ref{gen-e}) to leading
order in the finite-size expansion; note the cancellation of 
the first correction between numerator and denominator of
$\chi_\l$.

As is well known \cite{cft}, the partition functions of the rational CFTs,
such as (\ref{zucom}),
are invariant under modular transformations of the torus
made by the periodic space and compact time\cite{czmi};
moreover, the chiral parts $\chi_\l$ transform linearly among 
themselves \cite{cft}.
In particular, for the $S$ modular transformation, $\t \to -1/\t$,
one finds:
\bea
&& Z(-1/\t,-1/\overline{\t}) = Z(\t,\overline{\t})\ ,\nl
&& \chi_\l(-1/\t) = \frac{1}{\sqrt{p}} \sum_{\l'=1}^p\ 
\exp\left(i2\pi\frac{\l\l'}{p}\right)\ \chi_{\l'}(\t)\ .
\label{s-trans}
\eea
One can use this transformation to simplify the calculation
of (\ref{eps1}), by mapping $|q|\lesssim 1$ to 
$|\tilde{q}|=\exp(-2\pi/\I\t) \ll1$, such that the sum (product) in
$\chi(\tilde{q})_\l$ can be approximated by the leading term.
One easily obtains:
\beq
J_Q =  \frac{\pi}{12 \b^2} \left[ 
1+ O\left( \exp\left(-\frac{2\pi^2R}{v\b} \frac{1}{p}
\right)\right) \right]\ .
\label{eps12}
\eeq
We remark that: {\it i)} the leading term arises from the
prefactor $\tilde{q}^{-1/24}$ of the Dedekind function
in the characters (\ref{zucom}); 
{\it ii)} there are no power-law $O\left((v\b/R)^k\right)$
finite-size corrections\footnote{
One can check that the power-law corrections also vanish in the earlier 
expansion (\ref{epsc1}), but note that the Euler-Maclaurin formula
fails to reproduce the non-analytic terms in (\ref{eps12}).}.
The dominant term in the partition function for $|q| \to 0$ has
actually the general form $q^{-c/24}$ involving the Virasoro 
central charge, thus proving again the general
formula for the thermal conductance (\ref{gen-k}).
Furthermore, the derivation using the modular transformation
can be extended to any rational CFTs, that possesses a modular invariant 
partition function and a finite basis of characters linearly transforming
among themselves.
We conclude that all rational CFTs do not display any power-law 
finite-size correction to the thermal conductivity in (\ref{gen-k}).

%-3.2-------------------------------------
\subsection{Abelian conformal theories for the hierarchy}

We now consider the multi-component generalizations of
the scalar theory (also called multi-component Abelian CFT),
that could describe the hierarchical Hall states
with $\nu=m/(mp \pm 1)$:
for the plateaus at $\nu =2/(2p\pm 1)$, $p=2,4,\dots$,
such CFTs have central charge $c=2$, i.e. two propagating modes
per edge, one charged and one neutral \cite{abe}.
Although the chirality of the charged excitations is fixed by the external
magnetic field, neutral excitations could move in either 
direction (see Fig.2), leading to 
$(c_1,\bar{c}_1)=(2,0)$ or $(1,1)$ in the previous notation.
These theories are rational CFTs and their Abelian current algebra 
is extended to $\widehat{U(1)}\times \widehat{SU(2)}_{1}$;
the modular invariant partition functions are
again given by a finite sum of the chiral parts for 
each edge, that are themselves sums of representation-theory characters, 
as follows \cite{czmi}:
\beq
Z^{(\pm)}=\sum_{a=1}^{\widehat{p}}\ 
\theta^{(\pm)}_a \ \overline{\theta}^{(\pm)}_a\ ,
\label{zed2}
\eeq
where $+$ (resp. $-$) indicates propagation of the neutral mode
parallel (resp. antiparallel) to the charged one.
The chiral partition functions are:
\bea
\theta^{(+)}_a &=& \sum\limits_{\alpha =0}^{1}\ 
\chi _{2a+\alpha \widehat{p}}^{\widehat{U(1)}}\ 
\chi _{\alpha }^{\widehat{SU(2)_{1}}}\ ,\qquad a=1,2,\dots, \hat{p},\nl
\theta^{(-)}_a &=& \sum\limits_{\alpha =0}^{1}\ 
\chi _{2a+\alpha \widehat{p}}^{\widehat{U(1)}}\ 
\overline{\chi}_{\alpha }^{\widehat{SU(2)_{1}}}\ ,
\label{thetach}
\eea
where the characters are given by,
\beq
\chi _{\lambda }^{\widehat{U(1)}}(\t)\ =\ 
\frac{1}{\eta (\t)}\sum_{k=-\infty}^\infty\
q^{(2\hat{p}k+\lambda )^{2}/4\widehat{p}}\ ,
\qquad \chi _{\alpha}^{\widehat{SU(2)_{1}}}(\t)\ 
=\ \frac{1}{\eta (\t)}\sum_{k=-\infty}^\infty\ q^{(2k+\alpha )^{2}/4}\ .
\label{c2char}
\eeq
The $\u1$ and  $\su2$ characters describe the charged $(c)$ and 
neutral $(n)$ modes, respectively; $\alpha=0,1$ is the $\su2$ isospin parity, 
${\widehat{p}}=2/\nu=2p \pm 1$ and $\lambda =2a+\alpha \widehat{p}$
counts the units of fractional charge.
The two modes can have different velocities, which are accounted for 
by redefining the $q$ parameters in the corresponding characters:
$q \to q_j = \exp\left( -(\b -i \g_j)v_j/R \right)$, with $j=c,n$. 
Note that independent rescalings are possible because the Hamiltonian is
factorized: $H=H_c+H_n$ \cite{ham}.

In order to compute the thermal current on one edge (\ref{jq}), 
$J_Q=v_c \eps_c \pm v_n \eps_n $, we
vary the chiral partition functions (\ref{thetach}) as follows:
\beq
- \left. \frac{i}{2\pi R}
\frac{\partial \log\theta^{(\pm)}_\l}{\partial \g_c}
\right\vert_{\g=0} = \eps_c \ ,\qquad
- \left. \frac{i}{2\pi R}
\frac{\partial \log\theta^{(\pm)}_\l}{\partial \g_n}
\right\vert_{\g=0} = \pm\ \eps_n \ .
\label{eme}
\eeq
The computation of (\ref{eme}) to leading order in $(\b/R)$
can be  done as before, by performing a modular
transformation: the characters in (\ref{c2char}) and
chiral partition functions $\theta^{(\pm)}_a$ (\ref{thetach})
undergo finite Fourier transforms as in (\ref{s-trans}) \cite{czmi};
by expanding the resulting sums, the leading term reproduces the
thermal conductance in agreement with (\ref{gen-k}) and the subleading
terms are not polynomial in $1/R$.
We can actually give the general result for all the hierarchical
plateaus, as described by the multi-component scalar theories
(whose modular invariant partition functions can be found in Ref.\cite{czmi}).
These correspond to the $c=m$ CFTs with current algebra
$\widehat{U(1)}\times \widehat{SU(m)}_{1}$,
the second factor pertaining to the $(m-1)$ neutral modes with equal chirality,
namely $(c_1,\bar{c}_1)=(m,0)$ or $(1,m-1)$.
The result is:
\beq
K_{\rm Abelian} = \frac{\pi k_{B}^{2}T}{6}
\left[ 1\pm (m-1)\right]\ ,\qquad\quad{\rm for}\ \  \nu =\frac{m}{mp\pm 1}\ .
\label{jqabe}
\eeq

%--fig2-----------------------------------
\begin{figure}
%\centering
%\centering
\epsfysize=4cm
\centerline{\epsfbox{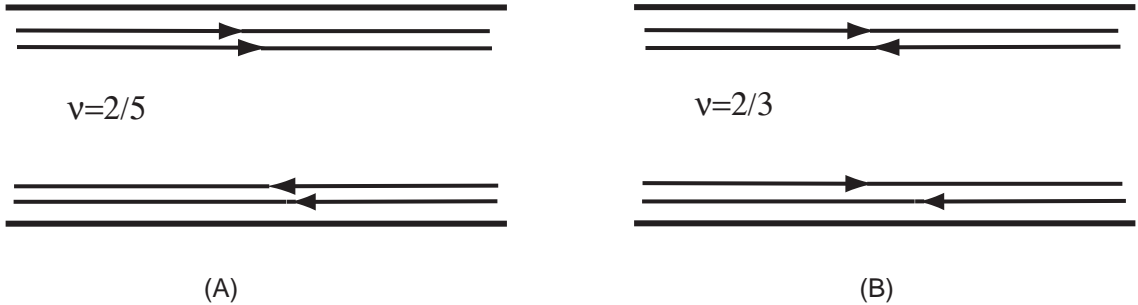}}
\caption{Neutral and charged modes in the plateaus with 
$\nu=2/5$ and $\nu =2/3$, having equal and opposite senses of propagation, 
respectively.}
\end{figure}
%---------------------------------------------

%-4--------------------------------------
\section{$\winf$ minimal models for the hierarchy}

We now focus our attention to another CFTs for
the hierarchical edge states, the $\winf$ minimal models,
which have been introduced in Ref.\cite{ctzmin} and further
analyzed in the Refs.\cite{czmi}\cite{ham}.
These models have been derived from the
requirement of the $\winf$ symmetry, which is the characteristic symmetry
of the incompressible fluids under area-preserving reparametrizations
of the spatial coordinates \cite{ctz}\cite{iks}. 
This symmetry can be naturally implemented 
in the CFTs describing the edge excitations of the Hall droplet. 
Another requirement is the minimality of the set of excitations, 
which translates into irreducibility of the $\winf$ representations \cite{kac}.
These two ingredients uniquely determine the $\winf$ minimal models,
that are actually in one-to-one correspondence 
with the Jain hierarchical plateaus \cite{ctzmin}.

The relation between the minimal and Abelian theories
has been discussed in Ref.\cite{ham}:
each minimal model can be obtained by projecting some neutral
states out of the Abelian theory for the same plateau.
In the $c=2$ case, the projection has been done very
explicitly: a term was added to the Abelian Hamiltonian,
parametrized by a positive coupling $\omega$, as follows:
\beq
H=\frac{1}{R}\left( v_c\ L_{0}^{(c)} + v_n\ L_{0}^{(n)} -
\frac{1}{12 } \right) + \omega J_{0}^{+} J_{0}^{-}\ . 
\label{hamin}
\eeq
This defines a theory interpolating between the Abelian ($\omega=0$) and
minimal ($\omega=\infty)$ model.
Actually, the neutral states in the Abelian theory carry
a $SU(2)$ isospin label, besides the Virasoro dimension,
that implies degenerate multiplets:
there are $(2s+1)$ Virasoro representations with conformal dimension $h=s^2$.
The additional term in the Hamiltonian (\ref{hamin}) assigns a large
weight $O(\omega)$ to all the states in the multiplets
except the highest weight state $|s,-s\rangle$ that satisfies 
$J^-_0 |s,-s\rangle =0$.
As a result, the $\winf$ minimal models have no multiplicities for the
Virasoro representations and the $SU(2)$ symmetry is broken.

The parameter $\omega$ has the dimension of a mass, thus the term
$J_{0}^{+} J_{0}^{-}$
is a relevant perturbation that also breaks conformal
invariance up to the infrared limit ($\omega=\infty$).
This renormalization-group flow takes place in the same phase of the
system, because the ground-state ($s=0$) is not affected by the projection.
The Hamiltonian (\ref{hamin}) naturally suggests the physical relevance of  
the $\winf$ minimal models: since the CFTs  are effective
low-energy, long-distance descriptions of the edge dynamics, the farthest
infrared fixed-point is physically relevant; in other words, $\omega \to
\infty $ is naturally reached without fine-tuning (if switched on).
The numerical energy spectrum
of 10 electrons in the first Landau level 
has been analyzed in Ref.\cite{cmsz} for the disk geometry: 
the low-energy levels
can be consistently interpreted by the edge excitations of the
$\winf$ minimal models, because the isospin multiplicities are not 
observed (up to the finite-size uncertainties).

The partition functions of the minimal models have the same 
structure as those of the Abelian theories (\ref{zed2}):
\beq
Z^{(\pm)}=\sum_{a=1}^{\widehat{p}}\ 
\Theta^{(\pm)}_a \ \overline{\Theta}^{(\pm)}_a\ ,
\label{zed2w}
\eeq
but the characters for the neutral sector are different.
The expressions for the $c=m=2$ case are \cite{czmi}:
\bea
\Theta^{(+)} &=& \sum\limits_{\alpha =0}^{1}\ 
\chi _{2a+\alpha \widehat{p}}^{\widehat{U(1)}}\ 
\chi _{\alpha }^{(n)}\ ,\nl
\Theta^{(-)} &=& \sum\limits_{\alpha =0}^{1}\ 
\chi _{2a+\alpha \widehat{p}}^{\widehat{U(1)}}\ 
\overline{\chi}_{\alpha }^{(n)}\ ,
\label{thetaw}
\eea
where the $\chi _{\alpha }^{(n)}$ are sums of the characters
$\chi^{\rm Vir}_h$ of the $c=1$ Virasoro 
degenerate representations with conformal weights $h=n^2/4$, $n=0,1,2,\dots$
\cite{cft}, 
\beq
\chi_\a^{(n)}= \sum_{\ell=0}^\infty\
\chi^{\rm Vir}_{(2\ell +\a)^2/4} = \frac{q^{\a/4}}{\eta(\tau)}\ .
\label{minchar}
\eeq

The projection (\ref{hamin}) can be seen at the level of partition functions:
the $\su2$ characters $\chi_\a^{\su2}$ of the Abelian
theory (\ref{c2char}) can be continuously connected to the $\chi^{(n)}_\a$
of the minimal theory (\ref{minchar}) by varying $\omega$ from zero
to infinity. 
The resulting partition function (\ref{zed2w}) is not modular invariant,
namely the $\winf$ minimal models are not rational CFTs.
The reason for this fact can again be traced back to the Hamiltonian
(\ref{hamin}): the $\omega$-perturbation is actually made by the non-local
operator $J^+_0 J^-_0$, thus the general arguments for modular invariance
in local field theory do not apply.
Note that the partition function (\ref{zed2w}) is uniquely
determined by the Hamiltonian definition of the minimal models
just outlined \cite{ham}.

We now proceed to compute the thermal conductance in the $c=2$
$\winf$ minimal models: using the definitions (\ref{eme}), 
we evaluate the average energies $\eps_c$ and $\eps_n$
following the same steps as in the Abelian case.
The $S$ modular transformation is still useful for expanding
the non-covariant neutral characters $\chi^{(n)}_\a$ in (\ref{minchar}):
we need the transformation of the Dedekind function,
\beq
\eta(\t)= (-i\t)^{-1/2}\ \eta\left( -1/\t \right)\ .
\label{s-dede}
\eeq
The result is (including the general case $c=m$):
\beq
K_{\rm Minimal} = \frac{\pi k_{B}^{2}T}{6}
\left[ 1\pm (m-1)\right]\ 
\mp\ \frac{k_B v_n}{4\pi R} (m-1)
\ ,\qquad\quad{\rm for}\ \ \ \nu =\frac{m}{mp\pm 1}\ .
\label{jqmin}
\eeq
This conductance contains an additional term w.r.t. the Abelian
result (\ref{jqabe}), namely the first finite-size correction $O(v\b/R)$.
One can check that it originates from the prefactor 
$(\I \t)^{-1/2}$  in the modular 
transformation of $\eta$ (\ref{s-dede}).

In summary, we have found that 
the thermal conductance in the $\winf$ minimal models (\ref{jqmin}) 
displays a finite-size correction to the general result 
(\ref{gen-k}) that is absent for all rational CFTs including
the Abelian theories of the hierarchical states.
Such correction is a consequence
of the lack of modular invariance in the minimal partition functions,
but can also be obtained  by straightforward expansion in $v\b/R $,
as described for the $c=1$ theory in Eq.(\ref{epsc1}); 
the latter derivation would make apparent that the correction arises from
the different state counting in the minimal models, which is
due to the projection of infinite states out of the Abelian theory.
It is also clear that this finite-size correction
in $K_{\rm Minimal}$ is universal and independent of the shape of the sample.

%-5----------------------------------
\section{Conclusions}

In this paper, we have found a general expression for the
thermal Hall conductance in terms of the 
gravitational anomaly of the chiral CFT describing each edge.
We have also obtained the finite-size expansion for the thermal 
conductance and computed the first universal, shape-independent
correction.
This was found to be different
for two candidate conformal theories of the hierarchical Hall states: the
multi-component Abelian theory has a vanishing correction, as any
rational CFT, while the $\winf$ minimal model has it non-vanishing.

The relative size of this correction w.r.t. the leading term 
$|(K_{\rm Minimal}-K_{\rm Abelian})/K|$
is equal to $x/2\pi\simeq \eps_{\rm edge}/\eps_{\rm therm}$,
with $x = v_n\beta/R$.
We estimate: $|\Delta K/K| \simeq 0.09 $ and $x\simeq 0.5$
for Hall samples of size $2\pi R\simeq 0.3\ cm$ 
and temperature $T\simeq 50\ mK$,  with $v_n \simeq c/1000$,
assuming comparable Fermi velocities for neutral and charged modes
\cite{ashoori}.
Although it is a small effect, it may be measurable in 
future experiments of the type proposed in \cite{kafi},
in the favorable cases, like $\nu=2/3$, where the leading term
in the thermal conductance is predicted to vanish. 
Such a difference could support either theoretical 
proposal for the hierarchical states.
Let us add that the CFT prediction for the thermal conductivity 
presented here is independent
of the impurity interactions, as discussed in Ref.\cite{kafi}.
Note also that further differences between the two CFTs for the hierarchical
theories involve the quantum statistics and require the measurement of
four-point functions.

Finally, the thermal Hall conductance can also be computed \cite{cri2}
for further conformal theories of the hierarchical states 
\cite{cri1} and for the paired Hall states \cite{more}.

{\bf Acknowledgments}

The work of GRZ is partially supported by grant PICT03-00000-00133
from ANPCYT, Argentina, and that of MH by 
C.N.R.S., France. AC acknowledges the partial 
support by the EC network grant FMRX-CT96-0012.
AC would like to thank the organizers of the workshop ``Strongly correlated
electron systems'', held in Capri, September 2001, for the
nice opportunity to present this work. 
MH thanks Horacio Casini for 
useful discussions. GRZ acknowledges the hospitality of the Abdus Salam 
Center for Theoretical Physics, Trieste, 
and the Physics Department and I.N.F.N. of Florence.

%------- Bibliography -------------------------------------
%
\def\NP{{\it Nucl. Phys.\ }}
\def\PRL{{\it Phys. Rev. Lett.\ }}
\def\PL{{\it Phys. Lett.\ }}
\def\PR{{\it Phys. Rev.\ }}
\def\CMP{{\it Comm. Math. Phys.\ }}
\def\IJMP{{\it Int. J. Mod. Phys.\ }}
\def\MPL{{\it Mod. Phys. Lett.\ }}
\def\RMP{{\it Rev. Mod. Phys.\ }}
\def\AP{{\it Ann. Phys.\ }}

\end{document}